\documentclass[
]{ceurart}

\usepackage{cleveref}
\usepackage{csquotes}
\usepackage{multirow}

\usepackage{listings}
\begin{document}

\copyrightyear{2021}
\copyrightclause{Copyright for this paper by its authors.
  Use permitted under Creative Commons License Attribution 4.0
  International (CC BY 4.0).}

\conference{INRA'21: 9th International Workshop on News Recommendation and Analytics, September 25, 2021, Amsterdam, Netherlands}

\title{How to Effectively Identify and Communicate Person-Targeting Media Bias in Daily News Consumption?}

\author[1,2,3]{Felix Hamborg}[%
orcid=0000-0003-2444-8056,
email=felix.hamborg@uni-konstanz.de,
url=https://felix.hamborg.eu/,
]
\address[1]{University of Konstanz, Konstanz, Germany}
\address[2]{Heidelberg Academy of Sciences and Humanities, Heidelberg, Germany}

\author[1]{Timo Spinde}[%
email=timo.spinde@uni-konstanz.de,
]

\author[1]{Kim Heinser}[%
email=kim.heinser@uni-konstanz.de,
]

\author[2,3]{Karsten Donnay}[%
orcid=0000-0002-9080-6539,
email=donnay@ipz.uzh.ch,
url=https://www.karstendonnay.net/,
]
\address[3]{University of Zurich, Zurich, Switzerland}

\author[2,4]{Bela Gipp}[%
orcid=0000-0001-6522-3019,
email=gipp@uni-wuppertal.de,
url=https://www.gipp.com/,
]
\address[4]{University of Wuppertal, Wuppertal, Germany}

\begin{abstract}
  Slanted news coverage strongly affects public opinion. This is especially true for coverage on politics and related issues, where studies have shown that bias in the news may influence elections and other collective decisions. Due to its viable importance, news coverage has long been studied in the social sciences, resulting in comprehensive models to describe it and effective yet costly methods to analyze it, such as content analysis. We present an in-progress system for news recommendation that is the first to automate the manual procedure of content analysis to reveal person-targeting biases in news articles reporting on policy issues. In a large-scale user study, we find very promising results regarding this interdisciplinary research direction. Our recommender detects and reveals substantial frames that are actually present in individual news articles. In contrast, prior work rather only facilitates the visibility of biases, e.g., by distinguishing left- and right-wing outlets. Further, our study shows that recommending news articles that differently frame an event significantly improves respondents' awareness of bias. 
\end{abstract}


\maketitle

\section{Introduction}\label{sec1}
How topics are covered in the news frames public debates and profoundly impacts collective decision-making, such as during elections \cite{dellavigna2006fox,Budak2019}. News may be subtly biased through various forms, such as word choice, framing, intentional omission or misrepresentation of specific details \cite{Hamborg2019d}. In extreme cases, ``fake news'' may present entirely fabricated facts to intentionally manipulate public opinion toward a given topic. A rich diversity of opinions is desirable, but systematically biased information can be problematic as a basis for decision-making if not recognized as such. Therefore, it is crucial to empower newsreaders in recognizing relative biases in coverage.

In this paper, we thus seek to answer the following research question. ``How can we effectively communicate instances of media bias in a set of news articles reporting on the same political event?'' Instead of identifying and then communicating each of the various forms of bias individually, we focus on person-oriented polarity, which is a fundamental effect resulting from various bias forms \cite{Hamborg2020a}. While the problem statement misses bias not related to persons, coverage on policy issues is largely person-oriented, e.g., because decisions are made by politicians or affect individuals in society. Our key contributions---especially when comparing to prior work---are: (1) We present the first system that is able to identify even subtle forms of media bias affecting the perception of persons by imitating the manual content analysis procedure from the social science. (2) We present modular visualizations to communicate media bias during daily news consumption. (3) We conduct a large scale user study using conjoint analysis to measure effectiveness of our analysis, visualizations, and individual components therein. 

We publish the survey materials, including questionnaires and anonymized answers, articles, and visualizations freely at: \url{https://doi.org/10.5281/zenodo.5517401}

\section{Related Work} 
\label{sec2}

Media bias has been long studied in the social sciences, resulting in a comprehensive set of models to describe it, such as political framing \cite{Entman2007a} and the news production process defining causes, forms, and effects of bias \cite{Hamborg2019d}, and effective methods to analyze it \cite{Hamborg2019d}. Established methods, such as content analysis and frame analysis \cite{Davis1975}, typically include systematic reading and labeling of texts. Despite their high effectiveness and reliability, they are largely conducted manually and do not scale with the vast amount of news. In contrast, many methods in computer science concerned with media bias employ automated and thus more efficient approaches but yield non-optimal results \cite{Hamborg2019d}, e.g., because they treat bias as only vaguely defined ``topic diversity'' \cite{park2011politics} or ``differences in [news] coverage'' \cite{Munson2009}. Though, methods for the identification of biased words exist for other domains, such as Wikipedia articles \cite{Recasens2013}.

Helping news consumers to become aware of media bias is an effective means to mitigate the negative effects of slanted news coverage, such as polarization \cite{Munson2009,mullainathan2005market}. Moreover, most studies find that automated approaches concerned with the communication of biases in the news can successfully increase bias-awareness in news consumers \cite{Kong2018,Park2009b,Park2011,Hamborg2020b}. However, previous approaches suffer from at least one of the following shortcomings. First, researchers cannot quantitatively pinpoint which individual components facilitate bias-awareness \cite{Kong2018,Park2009b,Park2011,Hamborg2020b}. Instead, studies measure overall effectiveness of analysis or visualizations. Second, approaches are concerned with the identification or communication of biases only in titles \cite{Kong2018}, on the article-level \cite{Munson2009,Hamborg2020b}, or outlet-level \cite{AllSides.com2021}. Considering only the overall article or even properties of its publisher, may lead to incorrect classification or missing instances of bias, e.g., that affect readers' perception on the sentence-level.

In sum, many approaches effectively communicate biases to users and most studies indicate how society benefits from doing so. However, many approaches identify only vaguely defined or superficial biases, e.g., because they do not use the established, effective models and analyses. Further, none of the reviewed approaches narrows down effectiveness regarding change in bias-awareness to individual analysis and visualization components. Lastly, to our knowledge, no approach identifies biases on the sentence level in news articles.

\section{System}
\label{sec3}

Given a set of news articles reporting on the same political event, our system's analysis aims to find groups of articles that frame the event similarly using four analysis tasks. These groups are later visualized (\Cref{sec4}) to enable non-expert news consumers to quickly get a bias-sensitive synopsis of a given news event.

For \textit{(1) article gathering}, we extract news articles reporting on one event \cite{Hamborg2017a}, currently for a set of user-defined URLs, or by providing texts to the system. We then perform state-of-the-art NLP \textit{(2) preprocessing} using Stanford CoreNLP. \textit{(3) Target concept analysis} finds and resolves person mentions across the topic's articles, including also broadly defined and event-specific coreferences that are otherwise non-coreferential or even opposing, such as ``freedom fighters'' and ``terrorists'' \cite{Hamborg2019d}. 

\textit{(4) Frame identification} determines how articles portray persons and then groups those articles that similarly portray (or frame) the persons. This task centers around political framing \cite{Entman2007a}, where a frame represents a specific perspective on an issue, e.g., which aspects are highlighted when reporting on the issue. While identifying frames would approximate content analyses as conducted in social science research on media bias more closely, it would yield lower classification performance \cite{Card2015,Card2016} or require infeasible effort since frames are typically created for a specific research-question \cite{Entman2007a}. Our system, however, is meant to analyze media bias caused by framing on any coverage reporting on policy issues. Thus, we seek to determine a fundamental effect resulting from framing: polarity of individual persons, which we identify on sentence- and aggregate to article-level. To achieve state of-the-art performance in target-dependent sentiment classification (TSC) on news articles, we use a fine-tuned RoBERTa-based neural model ($F1_m=83.1$) \cite{Hamborg2021b}.

The last step of frame identification is to determine groups of articles that similarly frame the event, i.e., the persons involved in the event. We currently use a simple, polarity-based method that first determines the person that occurs most frequently across all articles, named most frequent actor (\textit{MFA}). Then, the method assigns each article to one of three groups, depending on whether the article's MFA mentions are mostly positive, ambivalent, or negative. We also calculate each article's relevance to the (1) event and the (2) article's group using simple word-embedding scoring.

\section{Visualizations}
\label{sec4}
The overall workflow follows typical online news consumption, i.e., users see first an overview of news events and then view individual news articles. To measure effectiveness not only of our visualizations but also their constituents, we design them so that their components can be altered. To more precisely measure the change in bias-awareness concerning only the textual content, the visualizations show texts of articles (and information about biases in the texts) but no other content, e.g., photos and outlet name.

\subsection{Overview}
\label{sec4:overview}

The overview aims to enable users to quickly get a synopsis of a news event. We devise three visualizations. \textit{(1) Plain} represents popular news aggregators. Using a bias-agnostic design similar to Google News, this baseline shows article headlines and excerpts in a list sorted by their relevance to the event (\Cref{sec3}). \textit{(2) PolSides}, which represents a bias-aware news aggregator \cite{AllSides.com2021}, and \textit{(3) MFAP} share a bias-aware, comparative layout but use different methods to determine which frames or biases are present in event coverage. 

\begin{figure*}
  \includegraphics[width=\textwidth]{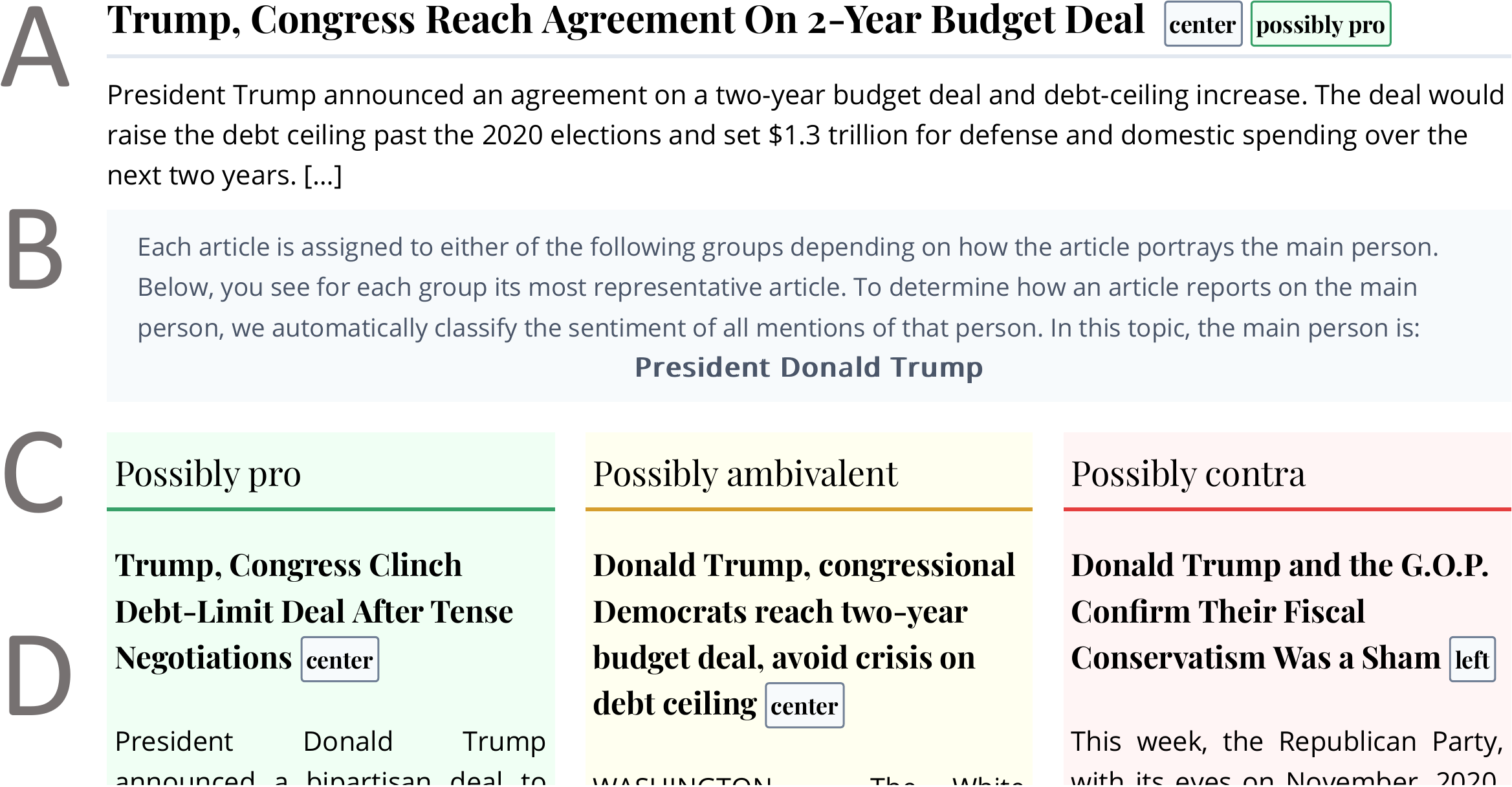}
  \caption{Excerpt of the MFAP overview showing three primary frames present in news coverage on a debt-ceiling event. 
  }
  \label{fig:teaser}
\end{figure*}

The layout of PolSides and MFAP is vertically divided in three parts, two of which are shown in \Cref{fig:teaser}. The event's \textit{main article} (part A) shows the event's most representative article. The comparative \textit{bias-groups} part (C) shows up to three frames present in event coverage, by showcasing each frame's most representative article. PolSides yields these frames by grouping articles depending on their political orientation (left, center, and right) \cite{AllSides.com2021}. For MFAP, we use our polarity-based grouping (\Cref{sec3}) so that the resulting groups represent frames that are primarily in favor, against, or ambivalent regarding the event's MFA. Conceptually, PolSides employs the left-right dichotomy, which is a simple yet often effective means to partition the media into distinctive slants. However, this dichotomy is determined only on the outlet-level and thus may incorrectly classify event-specific framing, e.g., articles with different perspectives having supposedly identical perspectives (and vice versa). Finally, a list shows the headlines of \textit{further articles} reporting on the event (bottom, not shown in \Cref{fig:teaser}).

In each overview, further components can be enabled depending on the conjoint profile (cf. \Cref{sec5}). \textit{PolSides tags} (shown close to D in \Cref{fig:teaser}) and/or \textit{MFAP tags} are shown next to each article headline, and indicate the political orientation of the article's outlet and the article's overall polarity regarding the MFA, respectively.

\subsection{Article View}
\label{sec4:articleview}
The article view shows an article's text and optionally the following visual clues to communicate bias information: \textit{(1) in-text polarity highlights}, \textit{(2) polarity context bar}, \textit{(3) PolSides tags}, and \textit{(4) MFAP tags} (with identical function to those in the overview). These clues are enabled, disabled, or altered depending on the conjoint profile. 

In-text polarity highlights aim to enable users to identify person-targeting sentiment on the sentence-level. We test the effectiveness of the following modes: \textit{single-color} (visually marking a person mention using a neutral color, i.e., gray, if the respective sentence mentions the person positively or negatively), \textit{two-color} (using green and red colors for positive and negative mentions, respectively), \textit{three-color} (same as two-color and additionally showing neutral polarity as gray), and \textit{disabled} (no highlights are shown).

\begin{figure}
    \centering
    \includegraphics[width=.7\linewidth]{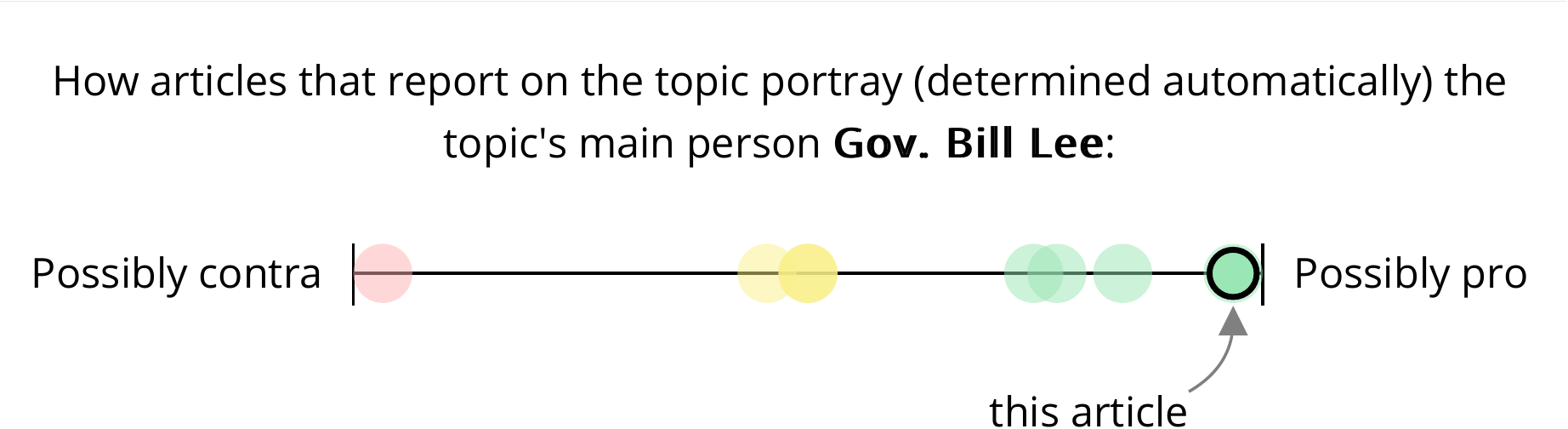}
    \caption{Polarity context bar showing the current and other articles polarity regarding the MFA.}
    \label{fig:polarity-context-bar}
\end{figure}

The \textit{polarity context bar} aims to enable users to quickly contrast how the current article and others portray the MFA. The 1D scatter plot depicted in \Cref{fig:polarity-context-bar} places articles as circles depending on their overall polarity regarding the MFA.  

\section{Experiments}
\label{sec5}
To evaluate the effectiveness of our system in supporting non-expert users to become aware of biases in news coverage, we conducted a user study consisting of two conjoint experiments. The study seeks to answers two research questions. \textit{What are effective means to communicate biases to non-expert news consumers when viewing an overview of a news topic (RQ1) and when reading a single news article (RQ2)?} The first experiment (E1) focuses on improving the general design of the overview (RQ1), while the second experiment (E2) focuses on answering both RQ1 with improved visualizations and RQ2. All survey data including questionnaires and anonymized respondents' information is available freely (\Cref{sec1}). 

\subsection{Methodology} 
\label{sec5:methodology}
In both experiments, we used a conjoint design to ``separately identify \textelp{} component-specific causal effects by randomly manipulating multiple attributes of alternatives simultaneously'' \cite{Hainmueller2014}. Respondents are asked to rate so-called ``profiles,'' which consist of multiple ``attributes,'' which are for example the overview, which topic it shows (or which article is shown in the article view), and if or which tags or in-text color highlights are shown. In conjoint design, these attributes are chosen randomly and independently of another for each respondent, which allows an estimation of the relative influence of each component on the bias-awareness (called Average Marginal Component Effects (AMCE)) \cite{Hainmueller2014}.

We selected three news topics to ensure varying degrees of expected polarization as an indicator for biased coverage: gun control (high polarization), debt ceiling (high-mid), and Australian bushfires (low). We selected a single event for each topic. To ensure heterogeneity in content and writing styles, we manually retrieved a balanced selection of ten articles from left-, center, and right-wing US online outlets as self-identified by them.

We conducted both experiments on Amazon Mechanical Turk. Respondents had to be located in the US, have a history of successfully completed, high quality work, and were compensated 1-2\$ depending on the study duration. In E1, we used data of 260 (of 308) respondents, which satisfied our quality measures, i.e., we discarded 48 respondents that, e.g., were unrealistically fast or answered test questions incorrectly. To keep cognitive load low, respondents were shown only a single topic in the overview, which was randomly drawn from the aforementioned selection. In E2, we used data of 98 (of 110) respondents. To increase cost efficiency, we only showed a selection of overview variants that exhibited positive trends in E1 (instead of fully randomly varying all attributes as in E1). Further, we showed respondents three tasks (resulting in 294 tasks in total), where each task consisted of a single overview and article view. 

Our study consists of seven steps. A \textit{(1) pre-study questionnaire} asks demographic data \cite{Spinde2020}. \textit{(2) Overview} (as described in \Cref{sec4:overview}). A \textit{(3) post-overview questionnaire} operationalizes the bias-awareness in respondents by asking about their perception of the diversity and disagreement in viewpoints, whether the visualization encourages contrasting the individual headlines, and how many perspectives of the public discourse were shown. While it is ``intrinsically difficult to objectively define what bias is'' \cite{Park2009b}, on a high level we expect bias to be \textit{perceived} in the form of differences in and opposition of the slant of articles; hence, we operationalize bias-awareness as the motivation and skill of a person to compare and contrast perspectives and information presented in the news using a set of 10-point Likert scaled questions, such as ``When shown the overview, did this encourage you to compare and contrast the different articles?''
\textit{(4) Article view} (as described in \Cref{sec4:articleview}). A \textit{(5) post-article questionnaire} operationalizes bias-awareness in respondents \cite{Spinde2020}. In a \textit{(6) post-study questionnaire}, users give feedback on the study, i.e., what they (dis)liked. E1 consisted of steps (1--3, 6). E2 consisted of all steps, where (2, 3) may be skipped depending on the conjoint profile and (2--5) were repeated three times since three topics were shown. 

\subsection{Results and Discussion}
We found positive, significant effects on the change in bias-awareness when using the overview variants PolSides or MFAP (RQ1) in E2. Tags also increased bias-awareness significantly. Regarding the article view (RQ2), E2 yielded insignificant results. 

Prior to E2, we focused our evaluation of E1 on qualitatively identifying flaws in the design of visualizations and the study, due to the lack of significant trends in E1.
For example, 35\% users experienced a lack of clarity and transparency, e.g., how the visualized information was derived. This weakness was exaggerated if respondents felt the shown information was incorrect (6\% PolSides, 11\% MFAP), e.g., an article that seemed negative from its headlines was labeled as ambivalent. 
Prior to E2, we addressed all the major lines of criticism, e.g., by adding brief explanations about all visual clues and in particular about the bias grouping (see B in \Cref{fig:teaser}). 

The goal of E2 was to test the set of overviews that had positive trends in E1 (to answer RQ1) and to test components in the article view (RQ2). E2 showed positive, significant effects of both bias-sensitive overviews, where the best overviews were PolSides (with PolSides tags) and MFAP (without tags). The AMCEs in \Cref{tab:effectiveness-e2} show that both overviews have very high and strongly significant effectiveness (PolSides $Est=7.83$ and MFAP $Est=6.13$, which are not significantly different to another albeit one being slightly higher \cite{Gelman2006}). Quantitative (by analyzing the effects on the individual questions composing the overall post-overview score) and qualitative analysis of both overviews suggests that MFAP reveals biases that are \textit{actually present} in the news articles, whereas PolSides \textit{only facilitates} the visibility of biases---a typical issue of prior approaches for automated bias detection \cite{Hamborg2020b}. By imitating the content analysis, our system yielded substantial frames as shown in \Cref{fig:teaser} whereas PolSides showed, e.g., the following headlines of rather ``artificial'' frames: ``Trump Announces Deal On Debt Limit, Spending Caps'' and ``Trump, Congress Clinch Debt-Limit Deal After Tense Negotiations.''
MFAP (random), an overview where articles were randomly assigned to one bias-group, yielded $Est=5.76$, indicating that only pointing out possible biases already increased bias-awareness. 

\begin{table}[]
\caption{Effects in E2 (excerpt). Names in parentheses indicate enabled tags (except for ``random''). Baselines are italic.}
\label{tab:effectiveness-e2}
\begin{tabular}{llrrrr}
\toprule
\textbf{Attribute} & \textbf{Level}       & \textbf{Est.} & \textbf{SE} & \textbf{z}  & \textbf{Pr(\textgreater{}|z|)} \\
\midrule
\multirow{10}{*}{Overview} & PolS. (PolS.)             & \textbf{7.83}  & 2.06  & 3.79  & <.001***         \\
 &MFAP (both)      & 5.87  & 1.73  & 3.39  & <.001***         \\
 &MFAP (none)        & \textbf{6.13}  & 2.00  & 3.05  & .002**          \\
 &MFAP (MFAP)      & 5.67  & 2.10  & 2.69  & .007**          \\
 &MFAP (PolS.)  & 5.60  & 1.85  & 3.02  & .002**          \\
 &MFAP (random)         & \textbf{5.76}  & 2.40  & 2.39  & .016*           \\
 &Plain (both)     & 3.66  & 1.90  & 1.92  & .053            \\
 &Plain (MFAP)     & 1.28  & 2.24  & 0.57  & .568            \\
 &Plain (PolS.) & 3.04  & 2.07  & 1.46  & .142            \\
 &\textit{Plain (none)} & 0  & -  & - &    -         \\
\midrule
\multirow{3}{*}{Topic} & Debt Ceiling    & -1.52 & 0.78  & -1.93 & .052            \\
 & Gun control         & -1.14 & 0.84  & -1.34 & .179     \\
 & \textit{Bushfire}         & 0 & -  & - & -     \\
\bottomrule
\end{tabular}

\end{table}

In MFAP, showing no tags yielded the highest effectiveness ($6.13$ compared to $5.87$ when both tags were shown). This indicates that the MFAP design reveals ``enough'' bias information and further visual clues may yield too complex visualizations. None of the tags alone have significant effects when combined with the Plain version, indicating that a bias-group layout is necessary for bias-awareness. 

In the article view, only showing the PolSides tags had a significant positive effect on bias awareness ($2.45$). There were no significant effects for the MFAP tags and polarity context bar. Analyzing respondents' criticism in the post-study questions, we attribute this to two shortcomings. First, too few in-text highlights to have a consistent effect (21\% of the article views had $\leq5$ highlights, 7\% had none). When controlling for the number of highlights, they had a significant, positive effect on bias-awareness. 
Second, there was a strong influence of individual topics on the effectiveness, e.g., respondents reported the debt ceiling topic was ``too complicated'' or ``boring'' to follow. We plan to address this by conducting a study with more respondents and a wider range of topics, to ensure a better representation of the public discourses.
Doing so will also strengthen the generalizability of the results and allow to investigate the effects of users' demographic data on their bias-awareness and change thereof
\cite{Coe2008}. Due to the small sample size in E2, our current analysis was inconclusive regarding demographic effects.

Although we did not filter for a representative sample of the US population, the distributions of our samples are approximately similar to the distributions of the US population in key dimensions such as age and political education.\footnote{For statistics of the sample please refer to the online resources, see \Cref{sec1}.} However, we propose to verify the generalizability of the study's findings to the entire US population or other countries using a larger respondent sample. Further, in E1 and E2 we assumed that MTurk workers are mostly non-expert news consumers. To verify this, we propose to explicitly ask for participants' degree of media literacy. 

\section{Conclusion}
We present the first system to automatically identify and then communicate person-targeting forms of bias in news articles reporting on policy events. Earlier, these biases could only be identified using content analyses, which--despite their effectiveness in capturing also subtle yet powerful biases--could only be conducted for few topics in the past due to their high cost, manual effort, and required expertise. In a large-scale user study, we employ a conjoint design to measure the effectiveness of visualizations and individual components. We find that our overviews significantly increase bias-awareness in respondents. In particular and in contrast to prior work, our bias-identification method seems to reveal biases that emerge from the content of news coverage and individual articles. In practical terms, our results suggest that the biases found and communicated by our method are actually \textit{present} in the news articles, whereas the reviewed prior work only \textit{facilitates} detection of biases, e.g., by distinguishing between left- and right-wing outlets. In sum, our exploratory work indicates the effectiveness of bias-sensitive news recommendation as a promising line of research for future work.

\begin{acknowledgments}
  This work is funded by the WIN program of the Heidelberg Academy of Sciences and Humanities, financed by the Ministry of Science, Research and the Arts of the State of Baden-Württemberg, Germany. The authors thank the anonymous reviewers for their valuable comments that helped to improve this paper. 
\end{acknowledgments}

\bibliography{refsfrozen}

\begin{thebibliography}{23}
\expandafter\ifx\csname natexlab\endcsname\relax\def\natexlab#1{#1}\fi
\providecommand{\url}[1]{\texttt{#1}}
\providecommand{\href}[2]{#2}
\providecommand{\path}[1]{#1}
\providecommand{\DOIprefix}{doi:}
\providecommand{\ArXivprefix}{arXiv:}
\providecommand{\URLprefix}{URL: }
\providecommand{\Pubmedprefix}{pmid:}
\providecommand{\doi}[1]{\href{http://dx.doi.org/#1}{\path{#1}}}
\providecommand{\Pubmed}[1]{\href{pmid:#1}{\path{#1}}}
\providecommand{\bibinfo}[2]{#2}
\ifx\xfnm\relax \def\xfnm[#1]{\unskip,\space#1}\fi
\bibitem[{DellaVigna and Kaplan(2006)}]{dellavigna2006fox}
\bibinfo{author}{S.~DellaVigna}, \bibinfo{author}{E.~Kaplan},
  \bibinfo{title}{{The Fox News Effect: Media Bias and Voting}},
  \bibinfo{type}{Technical Report} \bibinfo{number}{3}, National Bureau of
  Economic Research, \bibinfo{address}{Cambridge, MA}, \bibinfo{year}{2006}.
  \URLprefix \url{http://www.nber.org/papers/w12169.pdf}.
  \DOIprefix\doi{10.3386/w12169}.
\bibitem[{Budak(2019)}]{Budak2019}
\bibinfo{author}{C.~Budak},
\newblock \bibinfo{title}{{What happened? The Spread of Fake News Publisher
  Content During the 2016 U.S. Presidential Election}},
\newblock in: \bibinfo{booktitle}{The World Wide Web Conference on - WWW '19},
  \bibinfo{publisher}{ACM Press}, \bibinfo{address}{New York, New York, USA},
  \bibinfo{year}{2019}, pp. \bibinfo{pages}{139--150}. \URLprefix
  \url{http://dl.acm.org/citation.cfm?doid=3308558.3313721}.
  \DOIprefix\doi{10.1145/3308558.3313721}.
\bibitem[{Hamborg et~al.(2019)Hamborg, Donnay, and Gipp}]{Hamborg2019d}
\bibinfo{author}{F.~Hamborg}, \bibinfo{author}{K.~Donnay},
  \bibinfo{author}{B.~Gipp},
\newblock \bibinfo{title}{{Automated identification of media bias in news
  articles: an interdisciplinary literature review}},
\newblock \bibinfo{journal}{International Journal on Digital Libraries}
  \bibinfo{volume}{20} (\bibinfo{year}{2019}) \bibinfo{pages}{391--415}.
  \URLprefix \url{https://doi.org/10.1007/s00799-018-0261-y}.
  \DOIprefix\doi{10.1007/s00799-018-0261-y}.
\bibitem[{Hamborg(2020)}]{Hamborg2020a}
\bibinfo{author}{F.~Hamborg},
\newblock \bibinfo{title}{{Media Bias, the Social Sciences, and NLP: Automating
  Frame Analyses to Identify Bias by Word Choice and Labeling}},
\newblock in: \bibinfo{booktitle}{Proceedings of the 58th Annual Meeting of the
  Association for Computational Linguistics: Student Research Workshop},
  \bibinfo{publisher}{Association for Computational Linguistics},
  \bibinfo{address}{Stroudsburg, PA, USA}, \bibinfo{year}{2020}, pp.
  \bibinfo{pages}{79--87}. \URLprefix
  \url{https://www.aclweb.org/anthology/2020.acl-srw.12}.
  \DOIprefix\doi{10.18653/v1/2020.acl-srw.12}.
\bibitem[{Entman(2007)}]{Entman2007a}
\bibinfo{author}{R.~M. Entman},
\newblock \bibinfo{title}{{Framing Bias: Media in the Distribution of Power}},
\newblock \bibinfo{journal}{Journal of Communication} \bibinfo{volume}{57}
  (\bibinfo{year}{2007}) \bibinfo{pages}{163--173}. \URLprefix
  \url{https://academic.oup.com/joc/article/57/1/163-173/4102665}.
  \DOIprefix\doi{10.1111/j.1460-2466.2006.00336.x}.
\bibitem[{Davis and Goffman(1975)}]{Davis1975}
\bibinfo{author}{M.~S. Davis}, \bibinfo{author}{E.~Goffman},
\newblock \bibinfo{title}{{Frame Analysis: An Essay on the Organization of
  Experience.}},
\newblock \bibinfo{journal}{Contemporary Sociology} \bibinfo{volume}{4}
  (\bibinfo{year}{1975}) \bibinfo{pages}{599}. \URLprefix
  \url{http://www.jstor.org/stable/2064021?origin=crossref}.
  \DOIprefix\doi{10.2307/2064021}.
\bibitem[{Park et~al.(2011)Park, Ko, Kim, Liu, and Song}]{park2011politics}
\bibinfo{author}{S.~Park}, \bibinfo{author}{M.~Ko}, \bibinfo{author}{J.~Kim},
  \bibinfo{author}{Y.~Liu}, \bibinfo{author}{J.~Song},
\newblock \bibinfo{title}{{The politics of comments}},
\newblock in: \bibinfo{booktitle}{Proceedings of the ACM 2011 conference on
  Computer supported cooperative work - CSCW '11}, \bibinfo{organization}{ACM},
  \bibinfo{publisher}{ACM Press}, \bibinfo{address}{New York, New York, USA},
  \bibinfo{year}{2011}, p. \bibinfo{pages}{113}. \URLprefix
  \url{http://portal.acm.org/citation.cfm?doid=1958824.1958842}.
  \DOIprefix\doi{10.1145/1958824.1958842}.
\bibitem[{Munson et~al.(2009)Munson, Zhou, and Resnick}]{Munson2009}
\bibinfo{author}{S.~A. Munson}, \bibinfo{author}{D.~X. Zhou},
  \bibinfo{author}{P.~Resnick},
\newblock \bibinfo{title}{{Sidelines: An Algorithm for Increasing Diversity in
  News and Opinion Aggregators.}},
\newblock in: \bibinfo{booktitle}{ICWSM}, \bibinfo{year}{2009}.
\bibitem[{Recasens et~al.(2013)Recasens, Danescu-Niculescu-Mizil, and
  Jurafsky}]{Recasens2013}
\bibinfo{author}{M.~Recasens}, \bibinfo{author}{C.~Danescu-Niculescu-Mizil},
  \bibinfo{author}{D.~Jurafsky},
\newblock \bibinfo{title}{{Linguistic Models for Analyzing and Detecting Biased
  Language}},
\newblock in: \bibinfo{booktitle}{Proceedings of the 51st Annual Meeting on
  Association for Computational Linguistics}, \bibinfo{publisher}{Association
  for Computational Linguistics}, \bibinfo{address}{Sofia, BG},
  \bibinfo{year}{2013}, pp. \bibinfo{pages}{1650--1659}. \URLprefix
  \url{https://www.aclweb.org/anthology/P13-1162.pdf}.
\bibitem[{Mullainathan and Shleifer(2005)}]{mullainathan2005market}
\bibinfo{author}{S.~Mullainathan}, \bibinfo{author}{A.~Shleifer},
\newblock \bibinfo{title}{{The market for news}},
\newblock \bibinfo{journal}{American Economic Review}  (\bibinfo{year}{2005})
  \bibinfo{pages}{1031--1053}.
\bibitem[{Kong et~al.(2018)Kong, Liu, and Karahalios}]{Kong2018}
\bibinfo{author}{H.-K. Kong}, \bibinfo{author}{Z.~Liu},
  \bibinfo{author}{K.~Karahalios},
\newblock \bibinfo{title}{{Frames and Slants in Titles of Visualizations on
  Controversial Topics}},
\newblock in: \bibinfo{booktitle}{Proceedings of the 2018 CHI Conference on
  Human Factors in Computing Systems}, \bibinfo{publisher}{ACM},
  \bibinfo{address}{New York, NY, USA}, \bibinfo{year}{2018}, pp.
  \bibinfo{pages}{1--12}. \URLprefix
  \url{https://dl.acm.org/doi/10.1145/3173574.3174012}.
  \DOIprefix\doi{10.1145/3173574.3174012}.
\bibitem[{Park et~al.(2009)Park, Kang, Chung, and Song}]{Park2009b}
\bibinfo{author}{S.~Park}, \bibinfo{author}{S.~Kang},
  \bibinfo{author}{S.~Chung}, \bibinfo{author}{J.~Song},
\newblock \bibinfo{title}{{NewsCube: Delivering multiple aspects of news to
  mitigate media bias}},
\newblock in: \bibinfo{booktitle}{Proceedings of the 27th international
  conference on Human factors in computing systems - CHI 09},
  \bibinfo{publisher}{ACM Press}, \bibinfo{address}{New York, New York, USA},
  \bibinfo{year}{2009}, p. \bibinfo{pages}{443}. \URLprefix
  \url{http://dl.acm.org/citation.cfm?doid=1518701.1518772}.
  \DOIprefix\doi{10.1145/1518701.1518772}.
\bibitem[{Park et~al.(2011)Park, Ko, Kim, Choi, and Song}]{Park2011}
\bibinfo{author}{S.~Park}, \bibinfo{author}{M.~Ko}, \bibinfo{author}{J.~Kim},
  \bibinfo{author}{H.~Choi}, \bibinfo{author}{J.~Song},
\newblock \bibinfo{title}{{NewsCube 2.0: An Exploratory Design of a Social News
  Website for Media Bias Mitigation}},
\newblock in: \bibinfo{booktitle}{Workshop on Social Recommender Systems},
  \bibinfo{year}{2011}.
\bibitem[{Hamborg et~al.(2020)Hamborg, Meuschke, and Gipp}]{Hamborg2020b}
\bibinfo{author}{F.~Hamborg}, \bibinfo{author}{N.~Meuschke},
  \bibinfo{author}{B.~Gipp},
\newblock \bibinfo{title}{{Bias-aware news analysis using matrix-based news
  aggregation}},
\newblock \bibinfo{journal}{International Journal on Digital Libraries}
  \bibinfo{volume}{21} (\bibinfo{year}{2020}) \bibinfo{pages}{129--147}.
  \URLprefix \url{http://link.springer.com/10.1007/s00799-018-0239-9}.
  \DOIprefix\doi{10.1007/s00799-018-0239-9}.
\bibitem[{{AllSides.com}(2021)}]{AllSides.com2021}
\bibinfo{author}{{AllSides.com}}, \bibinfo{title}{{AllSides - balanced news}},
  \bibinfo{year}{2021}.
\bibitem[{Hamborg et~al.(2017)Hamborg, Meuschke, Breitinger, and
  Gipp}]{Hamborg2017a}
\bibinfo{author}{F.~Hamborg}, \bibinfo{author}{N.~Meuschke},
  \bibinfo{author}{C.~Breitinger}, \bibinfo{author}{B.~Gipp},
\newblock \bibinfo{title}{{news-please: A Generic News Crawler and Extractor}},
\newblock in: \bibinfo{booktitle}{Proceedings of the 15th International
  Symposium of Information Science}, \bibinfo{publisher}{Verlag Werner
  H{\"{u}}lsbusch}, \bibinfo{year}{2017}, pp. \bibinfo{pages}{218--223}.
\bibitem[{Card et~al.(2015)Card, Boydstun, Gross, Resnik, and Smith}]{Card2015}
\bibinfo{author}{D.~Card}, \bibinfo{author}{A.~E. Boydstun},
  \bibinfo{author}{J.~H. Gross}, \bibinfo{author}{P.~Resnik},
  \bibinfo{author}{N.~A. Smith},
\newblock \bibinfo{title}{{The Media Frames Corpus: Annotations of Frames
  Across Issues}},
\newblock in: \bibinfo{booktitle}{Proceedings of the 53rd Annual Meeting of the
  Association for Computational Linguistics and the 7th International Joint
  Conference on Natural Language Processing (Volume 2: Short Papers)},
  \bibinfo{publisher}{Association for Computational Linguistics},
  \bibinfo{address}{Stroudsburg, PA, USA}, \bibinfo{year}{2015}, pp.
  \bibinfo{pages}{438--444}. \URLprefix
  \url{http://aclweb.org/anthology/P15-2072}.
  \DOIprefix\doi{10.3115/v1/P15-2072}.
\bibitem[{Card et~al.(2016)Card, Gross, Boydstun, and Smith}]{Card2016}
\bibinfo{author}{D.~Card}, \bibinfo{author}{J.~Gross},
  \bibinfo{author}{A.~Boydstun}, \bibinfo{author}{N.~A. Smith},
\newblock \bibinfo{title}{{Analyzing Framing through the Casts of Characters in
  the News}},
\newblock in: \bibinfo{booktitle}{Proceedings of the 2016 Conference on
  Empirical Methods in Natural Language Processing},
  \bibinfo{publisher}{Association for Computational Linguistics},
  \bibinfo{address}{Stroudsburg, PA, USA}, \bibinfo{year}{2016}, pp.
  \bibinfo{pages}{1410--1420}. \URLprefix
  \url{http://aclweb.org/anthology/D16-1148}.
  \DOIprefix\doi{10.18653/v1/D16-1148}.
\bibitem[{Hamborg and Donnay(2021)}]{Hamborg2021b}
\bibinfo{author}{F.~Hamborg}, \bibinfo{author}{K.~Donnay},
\newblock \bibinfo{title}{Newsmtsc: (multi-)target-dependent sentiment
  classification in news articles},
\newblock in: \bibinfo{booktitle}{Proceedings of the 16th Conference of the
  European Chapter of the Association for Computational Linguistics (EACL
  2021)}, \bibinfo{year}{2021}, pp. \bibinfo{pages}{1663--1675}.
\bibitem[{Hainmueller et~al.(2014)Hainmueller, Hopkins, and
  Yamamoto}]{Hainmueller2014}
\bibinfo{author}{J.~Hainmueller}, \bibinfo{author}{D.~J. Hopkins},
  \bibinfo{author}{T.~Yamamoto},
\newblock \bibinfo{title}{{Causal Inference in Conjoint Analysis: Understanding
  Multidimensional Choices via Stated Preference Experiments}},
\newblock \bibinfo{journal}{Political Analysis} \bibinfo{volume}{22}
  (\bibinfo{year}{2014}) \bibinfo{pages}{1--30}. \URLprefix
  \url{https://www.cambridge.org/core/product/identifier/S1047198700013589/type/journal_article}.
  \DOIprefix\doi{10.1093/pan/mpt024}.
\bibitem[{Spinde et~al.(2020)Spinde, Hamborg, Donnay, Becerra, and
  Gipp}]{Spinde2020}
\bibinfo{author}{T.~Spinde}, \bibinfo{author}{F.~Hamborg},
  \bibinfo{author}{K.~Donnay}, \bibinfo{author}{A.~Becerra},
  \bibinfo{author}{B.~Gipp},
\newblock \bibinfo{title}{{Enabling News Consumers to View and Understand
  Biased News Coverage: A Study on the Perception and Visualization of Media
  Bias}},
\newblock in: \bibinfo{booktitle}{Proceedings of the ACM/IEEE Joint Conference
  on Digital Libraries in 2020}, \bibinfo{publisher}{ACM},
  \bibinfo{address}{New York, NY, USA}, \bibinfo{year}{2020}, pp.
  \bibinfo{pages}{389--392}. \URLprefix
  \url{https://dl.acm.org/doi/10.1145/3383583.3398619}.
  \DOIprefix\doi{10.1145/3383583.3398619}.
\bibitem[{Gelman and Stern(2006)}]{Gelman2006}
\bibinfo{author}{A.~Gelman}, \bibinfo{author}{H.~Stern},
\newblock \bibinfo{title}{{The Difference Between “Significant” and “Not
  Significant” is not Itself Statistically Significant}},
\newblock \bibinfo{journal}{The American Statistician} \bibinfo{volume}{60}
  (\bibinfo{year}{2006}) \bibinfo{pages}{328--331}. \URLprefix
  \url{http://www.tandfonline.com/doi/abs/10.1198/000313006X152649}.
  \DOIprefix\doi{10.1198/000313006X152649}.
\bibitem[{Coe et~al.(2008)Coe, Tewksbury, Bond, Drogos, Porter, Yahn, and
  Zhang}]{Coe2008}
\bibinfo{author}{K.~Coe}, \bibinfo{author}{D.~Tewksbury},
  \bibinfo{author}{B.~J. Bond}, \bibinfo{author}{K.~L. Drogos},
  \bibinfo{author}{R.~W. Porter}, \bibinfo{author}{A.~Yahn},
  \bibinfo{author}{Y.~Zhang},
\newblock \bibinfo{title}{{Hostile News: Partisan Use and Perceptions of Cable
  News Programming}},
\newblock \bibinfo{journal}{Journal of Communication} \bibinfo{volume}{58}
  (\bibinfo{year}{2008}) \bibinfo{pages}{201--219}. \URLprefix
  \url{https://academic.oup.com/joc/article/58/2/201-219/4098517}.
  \DOIprefix\doi{10.1111/j.1460-2466.2008.00381.x}.

\end{thebibliography}

\appendix

\end{document}